\documentclass[a4paper,11pt]{article}

\pdfoutput=1 

\usepackage{jinstpub} 
\usepackage{multirow}

\usepackage[version=4]{mhchem}
\usepackage{siunitx}
\usepackage{booktabs}
\usepackage[numbers]{natbib}
\usepackage{subcaption}

\title{Direct in-chamber radon-220 (thoron) emanation measurements for rare-event physics experiments}

\author[a,e,1]{R.R. Marcelo Gregorio,\note{Corresponding author.}}
\author[b]{F. Dastgiri,}
\author[b]{A. Basharina-Freshville,}
\author[a,e]{V.U. Bashu,}
\author[b]{A. Cottle,}
\author[a,e]{L. J. Bignell,}
\author[b]{C. Ghag,}
\author[a,e]{G. J. Lane,}
\author[a, c,e]{A.G. McLean}
\author[d]{and N.J.C. Spooner}

\affiliation[a]{ Department of Nuclear Physics and Accelerator Applications, Australian National University, \\Garran Road, ACT 2601, Canberra, Australia}
\affiliation[b]{Department of Physics and Astronomy, University College London, \\Gower Street, London, WC1E 6BT, United Kingdom}
\affiliation[c]{School of Physics, Chemistry and Earth Sciences, Adelaide University, \\North Terrace Campus, Adelaide, SA 5005, Australia }
\affiliation[d]{School of Mathematical and Physical Sciences, University of Sheffield, \\Hounsfield Road, Sheffield, South Yorkshire, S3 7RH, United Kingdom}
\affiliation[e]{ARC Centre of Excellence for Dark Matter Particle Physics, Australia}

\emailAdd{robert.gregorio@anu.edu.au}

\abstract{
Measuring radon emanation from detector materials is a key method for controlling radon, a significant background in rare-event physics experiments. Methods for measuring radon emanation are well-established but have predominantly focused on the $^{222}\mathrm{Rn}$ isotope, the dominant radon isotope for these backgrounds. However, measurements of $^{220}\mathrm{Rn}$ (thoron), the second most abundant radon isotope, remain relatively unexplored. ${}^{220}\mathrm{Rn}$ emanation measurements are challenging because the $^{220}\mathrm{Rn}$ must be transferred from the emanation chamber to the active detector within its short $55~\mathrm{s}$ half-life. In this study, a direct in-chamber approach for measuring ${}^{220}\mathrm{Rn}$ emanation is presented in which the sample is placed directly within the active detector chamber, thereby minimising losses during transfer. The method was demonstrated with a DURRIDGE RAD8 electrostatic radon detector, which measured ${}^{220}\mathrm{Rn}$ emanation from low-activity thoriated rods with an activity of $76 \pm 20~\text{mBq}$. Compared with a conventional flowthrough ${}^{220}\mathrm{Rn}$ emanation setup, the in-chamber method increased sensitivity by a factor of 3. Using helium as the carrier gas provided a further sensitivity increase, giving an overall sensitivity gain of $\sim$5. These results indicate that in-chamber ${}^{220}\mathrm{Rn}$ emanation measurements provide an effective tool for low-background experiments and have the potential to accelerate radon studies by exploiting the shorter half-life of $^{220}\mathrm{Rn}$.}

\keywords{Radon emanation, Radon-220, Thoron, Low-background techniques, Electrostatic collection }
\arxivnumber{1234.56789} 

\begin{document}

\maketitle
\flushbottom

\section{Introduction}

Radon contamination is a well-known source of unwanted background and remains a significant challenge for many rare-event physics experiments, such as dark matter and neutrinoless double beta decay searches~\cite{aalbers2025dark, bo2025dark, aprile2025wimp, dolinski2019neutrinoless}. The most significant isotope for radon-induced background is ${}^{222}\mathrm{Rn}$. The second most significant radon isotope, ${}^{220}\mathrm{Rn}$ (also known as thoron), has historically received less attention because of its half-life of 55.6 seconds which is short when compared with 3.8 days for ${}^{222}\mathrm{Rn}$. This short half-life limits the diffusion length of ${}^{220}\mathrm{Rn}$, suppressing the amount that can emanate from bulk materials into the active detector volume. However, ${}^{220}\mathrm{Rn}$ originating at or near detector surfaces can make a measurable contribution, and a ${}^{220}\mathrm{Rn}$ background component has been reported in several dark matter experiments~\cite{zc1w-88p6, marques20253d, amaudruz2015radon, Battat_2014}. 

One of the leading radon mitigation strategy is to perform extensive screening of detector materials by measuring their intrinsic radon emanation. Techniques for ${}^{222}\mathrm{Rn}$ emanation measurements are well-developed and achieve very high sensitivities using a range of established methods~\cite{ANKER2026170876, wu2025development, chott2023radon, zuzel2009high}. These approaches typically rely on accumulating ${}^{222}\mathrm{Rn}$ from an emanation chamber over long periods and then transferring the gas to a detector chamber. This procedure is poorly suited to ${}^{220}\mathrm{Rn}$ because of its short half-life. Although ${}^{220}\mathrm{Rn}$ monitoring instruments exist and are widely used in industrial applications, they are designed for comparatively high ${}^{220}\mathrm{Rn}$ levels ~\cite{nuccetelli1998thoron}, and their sensitivity is insufficient for the low activity levels required in rare-event physics. It is therefore important to develop dedicated ${}^{220}\mathrm{Rn}$ emanation techniques for low-background experiments.

Beyond its role in quantifying backgrounds, a dedicated ${}^{220}\mathrm{Rn}$ emanation facility provides a powerful diagnostic tool for material studies. In many rare-event detectors, particularly those operated at cryogenic temperatures, the dominant ${}^{222}\mathrm{Rn}$ contribution comes from recoil-induced emanation from surfaces and near-surface regions of detector components, rather than from diffusion out of bulk materials~\cite{perry2024background}. In this context, ${}^{220}\mathrm{Rn}$ can serve as a rapid first-pass proxy for ${}^{222}\mathrm{Rn}$ surface behaviour. 
This makes ${}^{220}\mathrm{Rn}$ emanation measurements a useful tool for supporting and accelerating studies of surface treatments and coatings designed to suppress radon emanation.

In this paper, the mechanisms for radon emanation measurements, with a focus on ${}^{220}\mathrm{Rn}$ and their limitations, are discussed in \autoref{sec:measurementofthoron}. To provide a baseline, the conventional ${}^{220}\mathrm{Rn}$ emanation method, its experimental setup, and subsequent analysis are described in \autoref{sec:experimentalstandard}. The improved in-chamber method, together with its experimental setup and analysis, is presented in \autoref{sec:inchamber}. A further enhancement using helium as the carrier gas is introduced in \autoref{sec:helium}. Finally, practical applications for low-background studies are discussed in \autoref{sec:application}.

\section{Measurement of $^{220}$Rn emanation}
\label{sec:measurementofthoron}

Radon emanation measurements typically proceed in three stages: an emanation period in a sealed chamber containing the sample of interest, transfer of the accumulated radon into a detection chamber, and subsequent measurement with a radon detector. The emanation period is required to allow radon to be produced from its parent radionuclides, escape from the sample, and accumulate in the chamber.
The emanation period is usually chosen so that the activity in the chamber is representative of the secular equilibrium value for the isotope of interest. In practice, this is achieved by waiting for several half-lives of the radon isotope. After about five to seven half-lives the activity has risen to within a few per cent (approximately 97–99\%) of its asymptotic value. For $^{220}\mathrm{Rn}$ (thoron), which has a half-life of 55 seconds, this corresponds to about 5–10 minutes of emanation with $^{224}\mathrm{Ra}$ as the parent. In contrast, $^{222}\mathrm{Rn}$, with a half-life of 3.8 days, typically requires 25–30 days to approach secular equilibrium with $^{226}\mathrm{Ra}$. Typically, shorter emanation times are used and the  accounted for in the analysis~\cite{Marcelo_Gregorio_2021, ogawa2024measurement}.

Once the emanation period is complete, the gas is transferred from the emanation chamber to the detection chamber. Many approaches exist for transferring radon between chambers, including complex setups based on physisorption on charcoal traps or liquefaction of radon~\cite{wiebe2024high, wu2025development}. While these methods work well for $^{222}\mathrm{Rn}$, they are poorly suited to $^{220}\mathrm{Rn}$ because the additional transfer steps introduce delays. As a result, practical ${}^{220}\mathrm{Rn}$ emanation measurements tend to rely on simple gas-flow transfer with short lines between the emanation and detector volumes.

\begin{figure}[h]
\centering
\includegraphics[height=6cm]{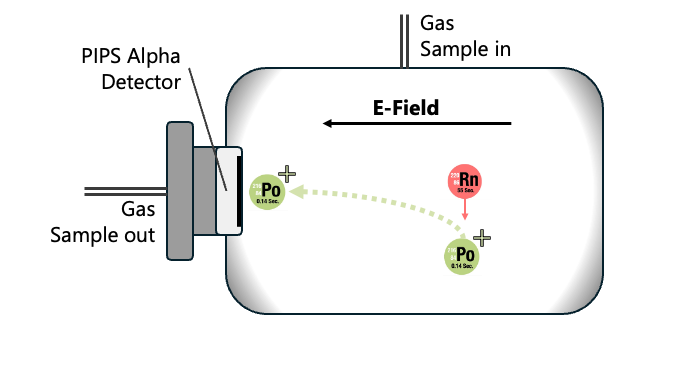}
\caption{Schematic of an electrostatic chamber, showing the mechanism of $^{220}\mathrm{Rn}$  detection.}
\label{fig:electrostatic_mech}
\end{figure}

Once the radon is inside the detection chamber, most emanation setups use electrostatic collection, which offers good energy resolution and can be adapted to different carrier gases. \autoref{fig:electrostatic_mech} shows a schematic of a typical electrostatic detector chamber. A high positive voltage is applied to the chamber walls, while the passivated implanted planar silicon (PIPS) detector is held at ground potential, generating an electric field within the active volume. ${}^{220}\mathrm{Rn}$ atoms that have been successfully transferred into the detector (red in \autoref{fig:electrostatic_mech}) decay to short-lived, positively charged progeny (green), which drift along the field lines to the PIPS active alpha detector surface before decaying. Their subsequent alpha decays are measured and converted to a ${}^{220}\mathrm{Rn}$ activity.

\begin{figure}[h]
\centering
\includegraphics[width=5.5cm]{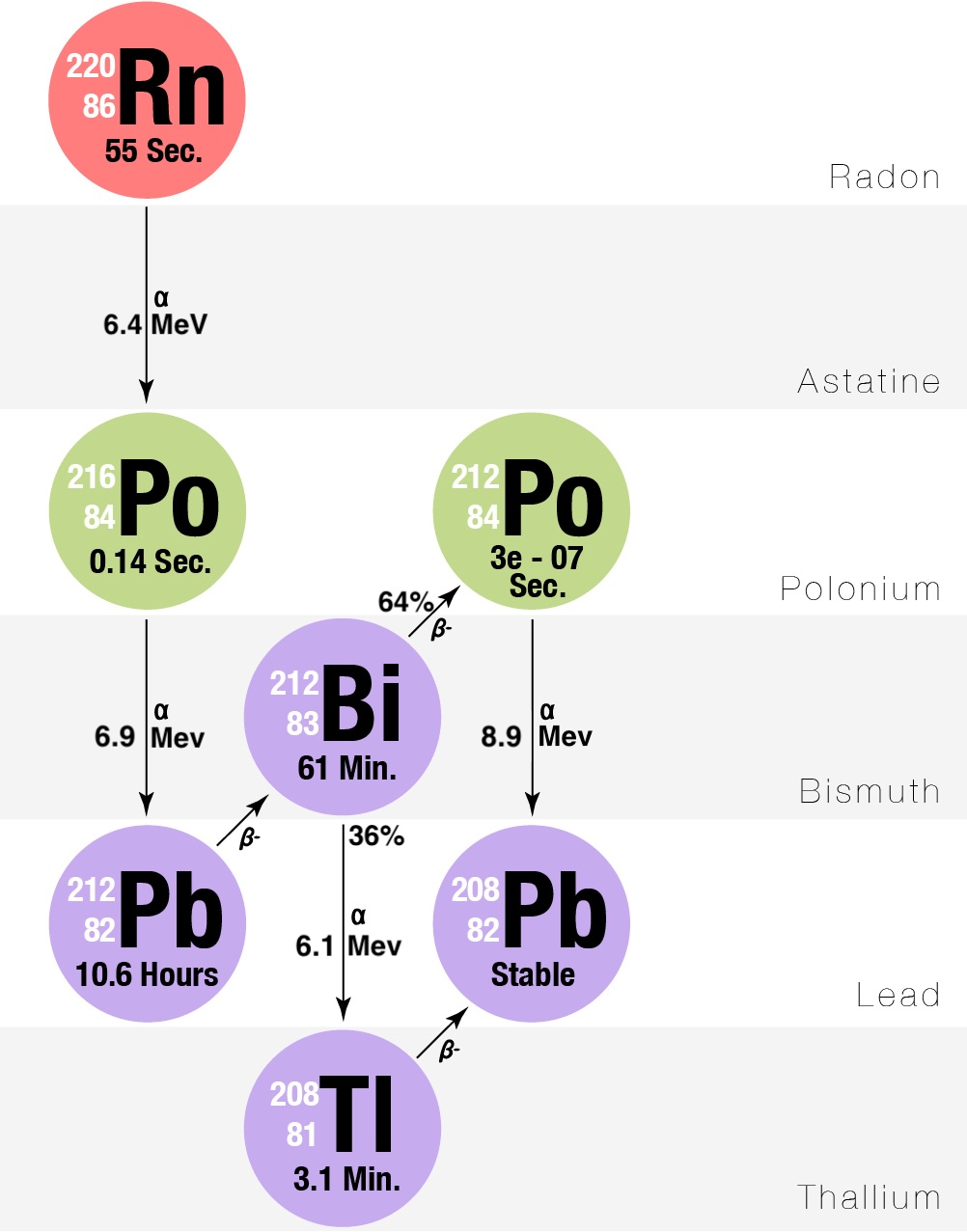}
\caption{Decay chain of ${}^{220}\mathrm{Rn}$ and its decay modes. Image adapted from~\cite{internachiThoronDecayChain}.}
\label{fig:thoron_chains}
\end{figure}

\autoref{fig:thoron_chains} shows the ${}^{220}\mathrm{Rn}$ decay chain, where information inside the circles indicate the nuclides and their half-lives, and the arrows indicates decay modes and alpha energies. For electrostatic ${}^{220}\mathrm{Rn}$ measurements, the primary collected progeny is ${}^{216}\mathrm{Po}$, which emits a 6.9~MeV alpha particle. Subsequent alpha decays from ${}^{212}\mathrm{Bi}$ (6.2~MeV, with an $\sim 36\%$ alpha-branching ratio) and ${}^{212}\mathrm{Po}$ (8.9~MeV) also appear in the spectrum. In practice, the activity is usually determined from the ${}^{216}\mathrm{Po}$ peak only, which responds almost immediately, whereas the ${}^{212}\mathrm{Bi}$ and ${}^{212}\mathrm{Po}$ peaks require of order 12~hours to reach their equilibrium intensities.

Although ${}^{220}\mathrm{Rn}$ is straightforward to detect once it reaches the detector chamber, providing distinct energy peaks that form an unambiguous signature, its very short half-life makes rapid transfer essential. To avoid decay losses, minimising the transfer time is crucial for accurate low-background ${}^{220}\mathrm{Rn}$ measurements.
Most conventional thoron measurement setups used in industrial applications account for these unavoidable losses by applying calibration factors, which effectively raises the lower limit of detection. This conventional approach is described in \autoref{sec:experimentalstandard}, and provides a baseline for comparison with the methods developed in this work.

\section{Flowthrough method}
\label{sec:experimentalstandard}

\begin{figure}[h]
\centering
\includegraphics[height=6cm]{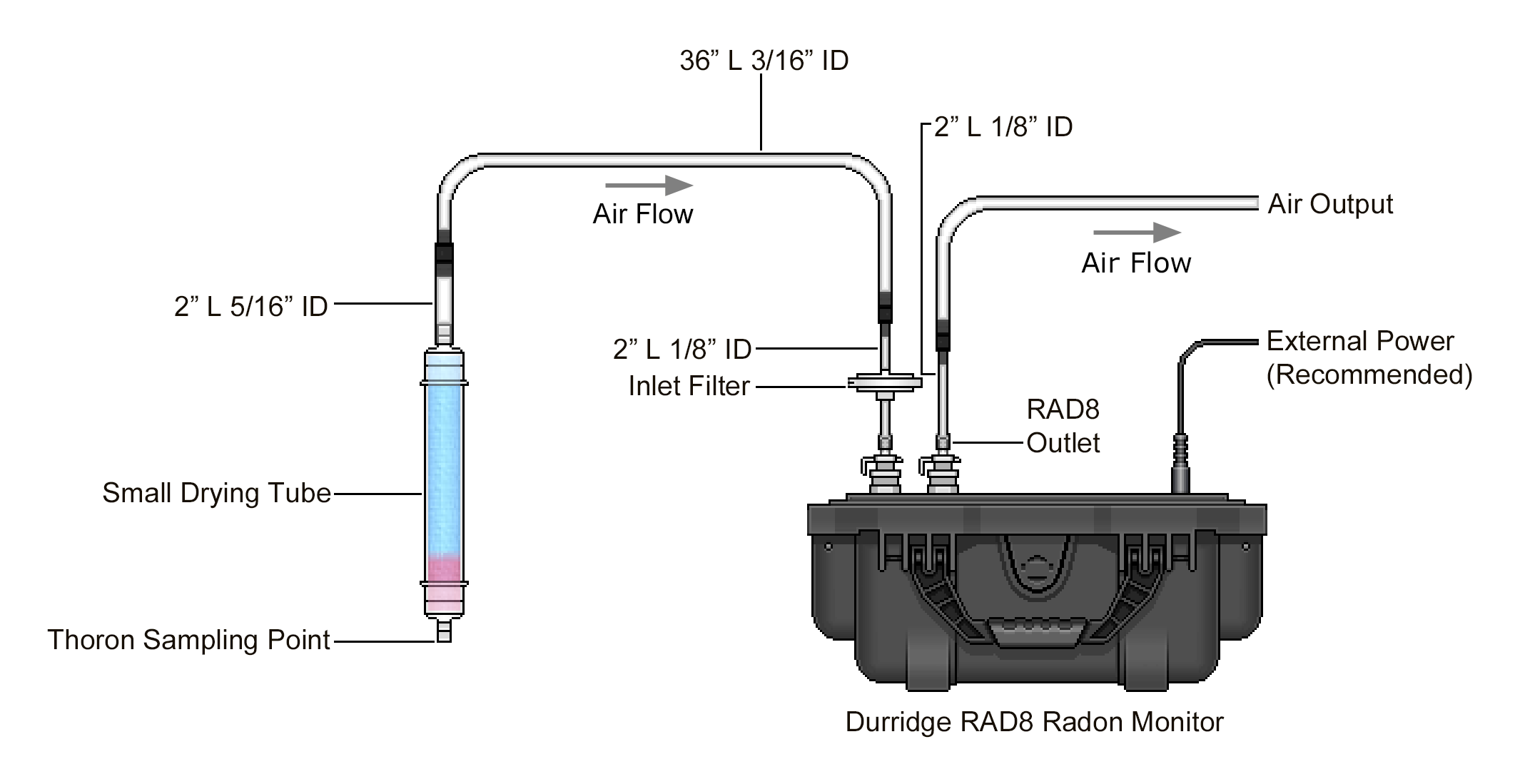}
\caption{DURRIDGE standard thoron flowthrough configuration. Image from~\cite{sadler_thoron_calibration_2024}.}
\label{fig:thoronstandard}
\end{figure}

\autoref{fig:thoronstandard} shows the DURRIDGE RAD8 standard thoron flowthrough configuration \cite{sadler_thoron_calibration_2024}. The setup consists of a thoron sampling point connected to a small drying tube filled with Drierite desiccant, followed by vinyl tubing and a \(0.45~\text{\textmu m}\) nylon inlet filter leading to the RAD8 detector. An internal diaphragm pump, with a nominal flow of \(0.6~\text{L min}^{-1}\), draws gas into the RAD8 electrostatic detector chamber. It is important that the tubing lengths and diameters, and the pump flow setting, were identical to those specified in the standard configuration, since transfer losses are accounted for in the factory calibration.

\begin{figure}[h]
\centering

\includegraphics[height=5cm]{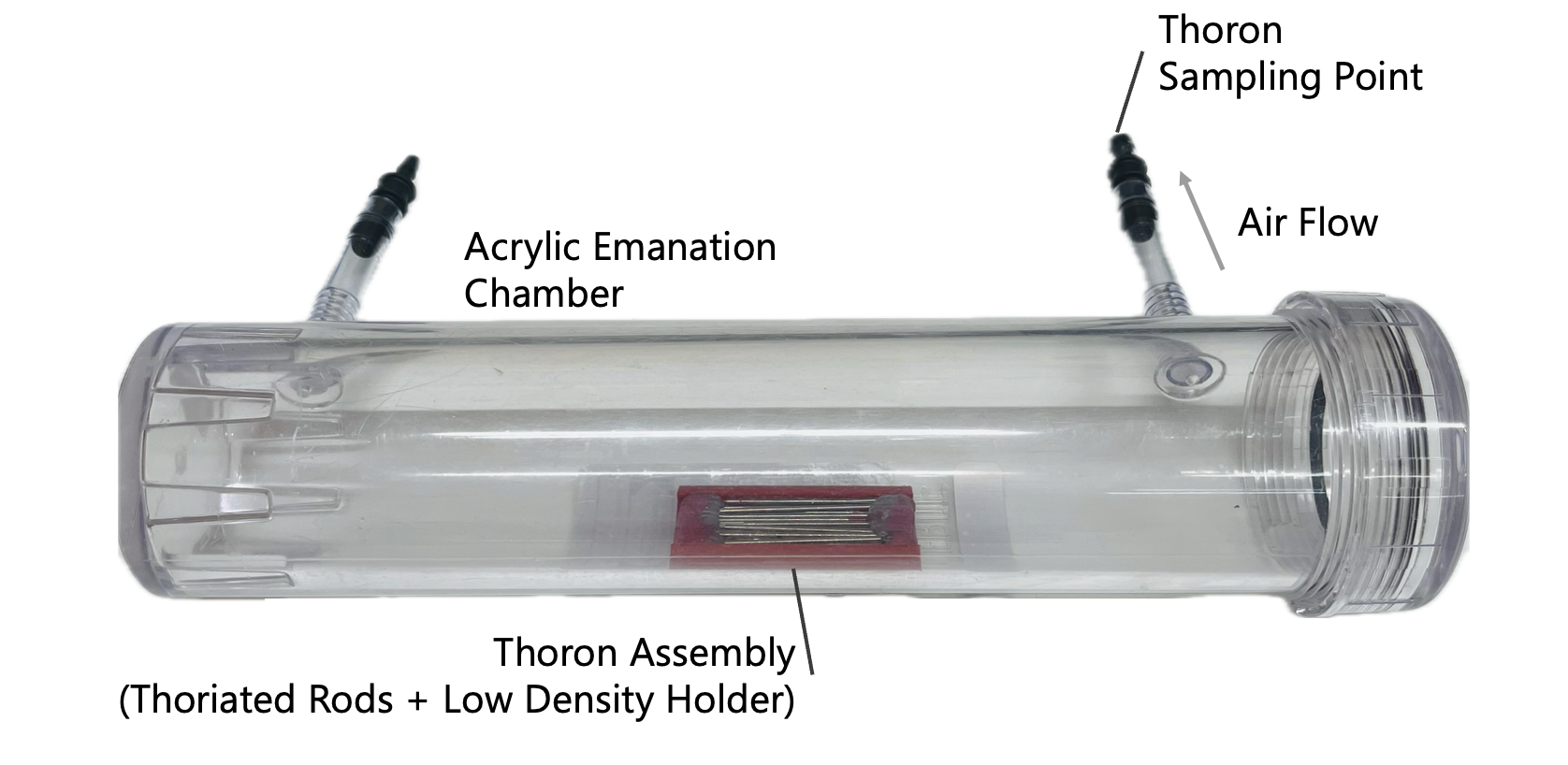}
\caption{Low-activity thoron source assembly inside the acrylic emanation chamber.}
\label{fig:weildinrods}
\end{figure}

\autoref{fig:weildinrods} shows the low-activity thoron source used in this study inside the acrylic emanation chamber. Thoriated tungsten rods been shown to demonstrated to provide a stable and convenient low-activity thoron source~\cite{wu2025development}. The thoron assembly consists of 31$~$g of 2\% thoriated tungsten welding rods with 1.6~mm diameter, secured in a low-density 3D printed holder. The outlet of the emanation chamber is connected to the thoron sampling point in the RAD8 standard thoron flowthrough setup.

\subsection*{Method}
The measurements used the RAD8 configuration and sampling protocol recommended by DURRIDGE as a baseline reference~\cite{sadler_thoron_calibration_2024}. Humidity affects the electrostatic collection efficiency, and the RAD8 operating guidelines recommend that the chamber relative humidity be maintained at $\leq 15\%$~\cite{instrumentation2017rad7}. The humidity was monitored using both the RAD8 internal humidity sensor and a small Drierite indicator, which changes from blue to pink on saturation. Measurement runs were carried out for 3 hours to maximise counting statistics while remaining within this recommended low-humidity operating range. Measurements were performed with the thoron assembly inside the acrylic emanation chamber and with the chamber empty, providing source and background data respectively.

\subsection*{Data analysis and results}
The RAD8 calculates radon concentrations using calibrated energy windows corresponding to alpha decays from both $^{222}\mathrm{Rn}$ and $^{220}\mathrm{Rn}$. For ${}^{220}\mathrm{Rn}$, the relevant alpha decays are $^{216}\mathrm{Po}$ (6.9~MeV), $^{212}\mathrm{Bi}$ (6.1~MeV), and $^{212}\mathrm{Po}$ (8.9~MeV). For $^{222}\mathrm{Rn}$, the dominant peaks are $^{218}\mathrm{Po}$ (6.00~MeV) and $^{214}\mathrm{Po}$ (7.69~MeV). The calibrated energy windows are denoted A, B, C and D, following the historical nomenclature used for radon progeny decays~\cite{rutherford2014radium}. Window A receives contributions from $^{218}\mathrm{Po}$ and $^{212}\mathrm{Bi}$, Window B is dominated by $^{216}\mathrm{Po}$, Window C by $^{214}\mathrm{Po}$, and Window D by $^{212}\mathrm{Po}$.

\begin{figure}[h]
  \centering
  \includegraphics[height=6.5cm]{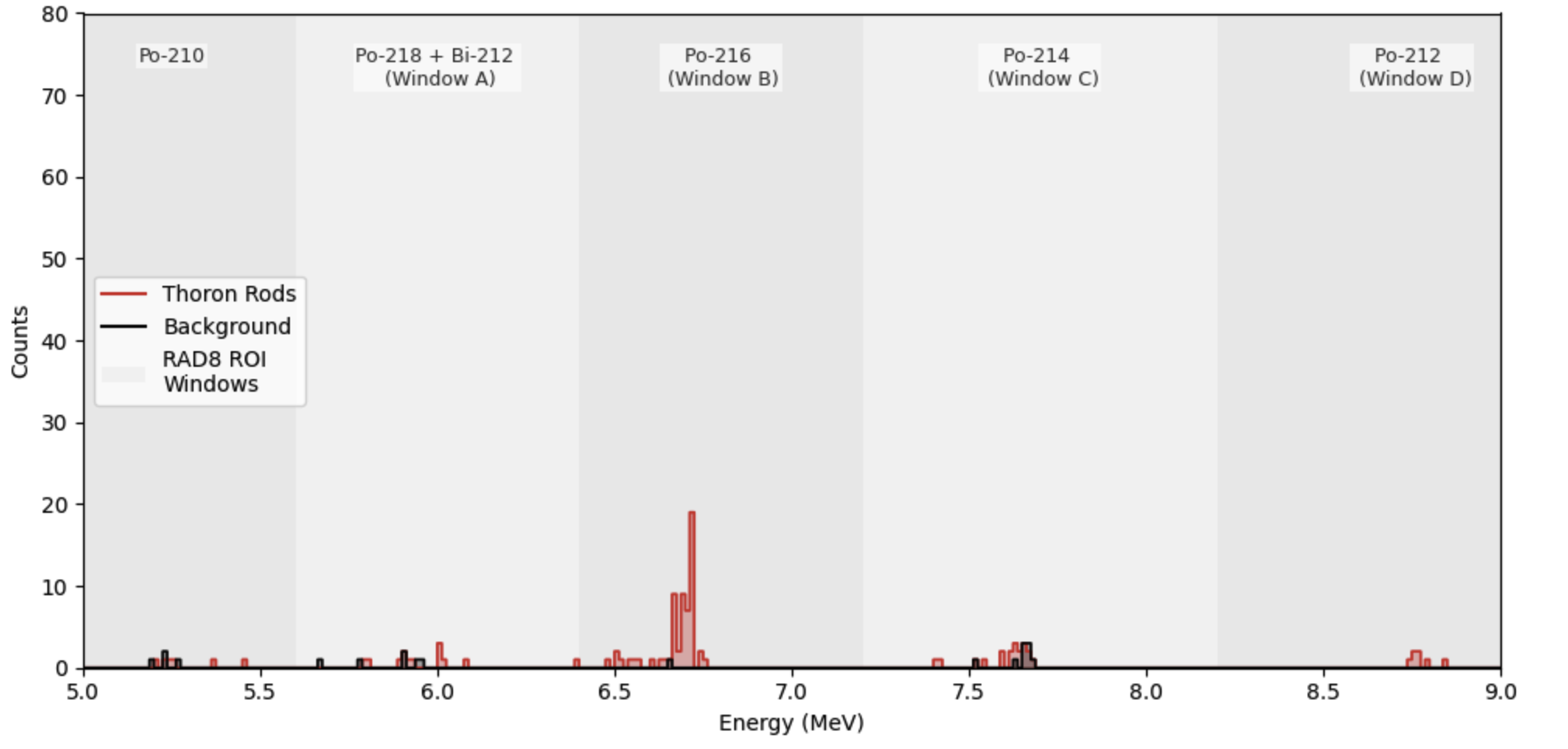}
  \caption{Alpha energy spectra for the thoron source (red) and empty emanation chamber background (black), measured with the flowthrough method over a 3~hour run, with the RAD8 calibrated energy windows A, B, C and D indicated.}
  \label{fig:flowthrough_espec}
\end{figure}

\autoref{fig:flowthrough_espec} shows the alpha energy spectrum integrated over a 3 hour run for the thoron assembly and for an empty emanation chamber measurement using the flowthrough method, shown in red and black, respectively. The relative humidity of the setup remained below 15\% throughout. 
A small excess above background is observed in Window B, consistent with $^{216}\mathrm{Po}$ from ${}^{220}\mathrm{Rn}$. A small excess is also visible in Window D, consistent with $^{212}\mathrm{Po}$ from later ${}^{220}\mathrm{Rn}$ progeny. Window A is difficult to interpret due to the overlap of $^{212}\mathrm{Bi}$ with $^{218}\mathrm{Po}$, while counts in Window C indicate a contribution from $^{222}\mathrm{Rn}$ originating from ambient radon in air.

To calculate the ${}^{220}\mathrm{Rn}$ concentration, only the count rate from $^{216}\mathrm{Po}$ in Window B is used. The short half-lives of $^{220}\mathrm{Rn}$ and $^{216}\mathrm{Po}$ mean that the $^{216}\mathrm{Po}$ activity reaches practical equilibrium within about 5~min of establishing flow and can be taken as ${}^{220}\mathrm{Rn}$ emanation rate.
The other windows are less suitable for quantification. Window A has overlapping contributions and is not specific to ${}^{220}\mathrm{Rn}$, and Window D is dominated by $^{212}\mathrm{Po}$, which does not reach equilibrium within a 3 hour run. Using the RAD8 thoron calibration in air, the background-subtracted Window~B rate for the thoriated rod assembly corresponds to a ${}^{220}\mathrm{Rn}$ concentration of $(94 \pm 25)~\mathrm{Bq\,m^{-3}}$.
To convert the measured ${}^{220}\mathrm{Rn}$ concentration to an activity $A_{220}$, the DURRIDGE thoron calibration relation was used~\cite{sadler_thoron_calibration_2024}.
\begin{equation}
  A_{220} \;=\; \frac{C_{220}\,Q}{\lambda_{220}}\,,
  \label{eq:th_activity}
\end{equation}
Here $C_{220}$ is the measured ${}^{220}\mathrm{Rn}$ concentration, $Q$ is the volumetric flow rate, and $\lambda_{220}$ is the $^{220}\mathrm{Rn}$ decay constant.  The ${}^{220}\mathrm{Rn}$ activity concentration in the air in the thoron assembly chamber was found to be $A_{220} = 76 \pm 20~\mathrm{mBq}$ (95\% C.L.). This result is used as the baseline for comparisons with the in-chamber method.

\section{In-chamber method}
\label{sec:inchamber}

\begin{figure}[h]
\centering
\includegraphics[height=6.5cm]{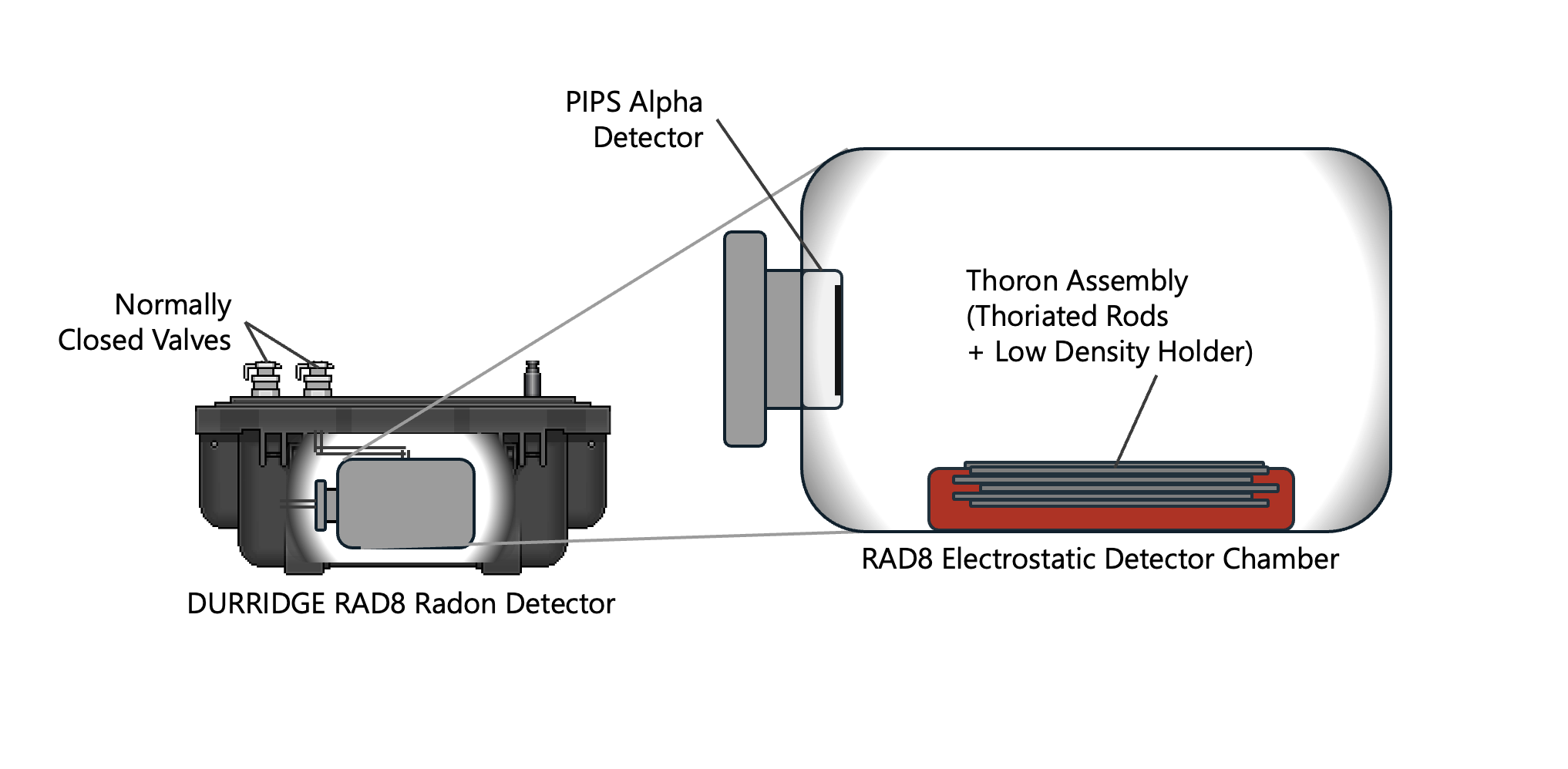} 
\caption{In-chamber configuration with thoron assembly inside the RAD8 detector chamber.}
\label{fig:RAD8MECH_internal}
\end{figure}

The in-chamber approach aims to improve sensitivity to $^{220}\mathrm{Rn}$ by minimising transfer losses between the emanation chamber and the detector chamber. \autoref{fig:RAD8MECH_internal} shows the in-chamber setup, where the same low-activity thoron source assembly, described in \autoref{sec:experimentalstandard}, is placed directly inside the RAD8 electrostatic detector chamber. The 3D printed holder used in the assembly provides electrical isolation from the chamber walls.

\subsection*{Method}
\begin{figure}[h]
\centering
\includegraphics[height=6.5cm]{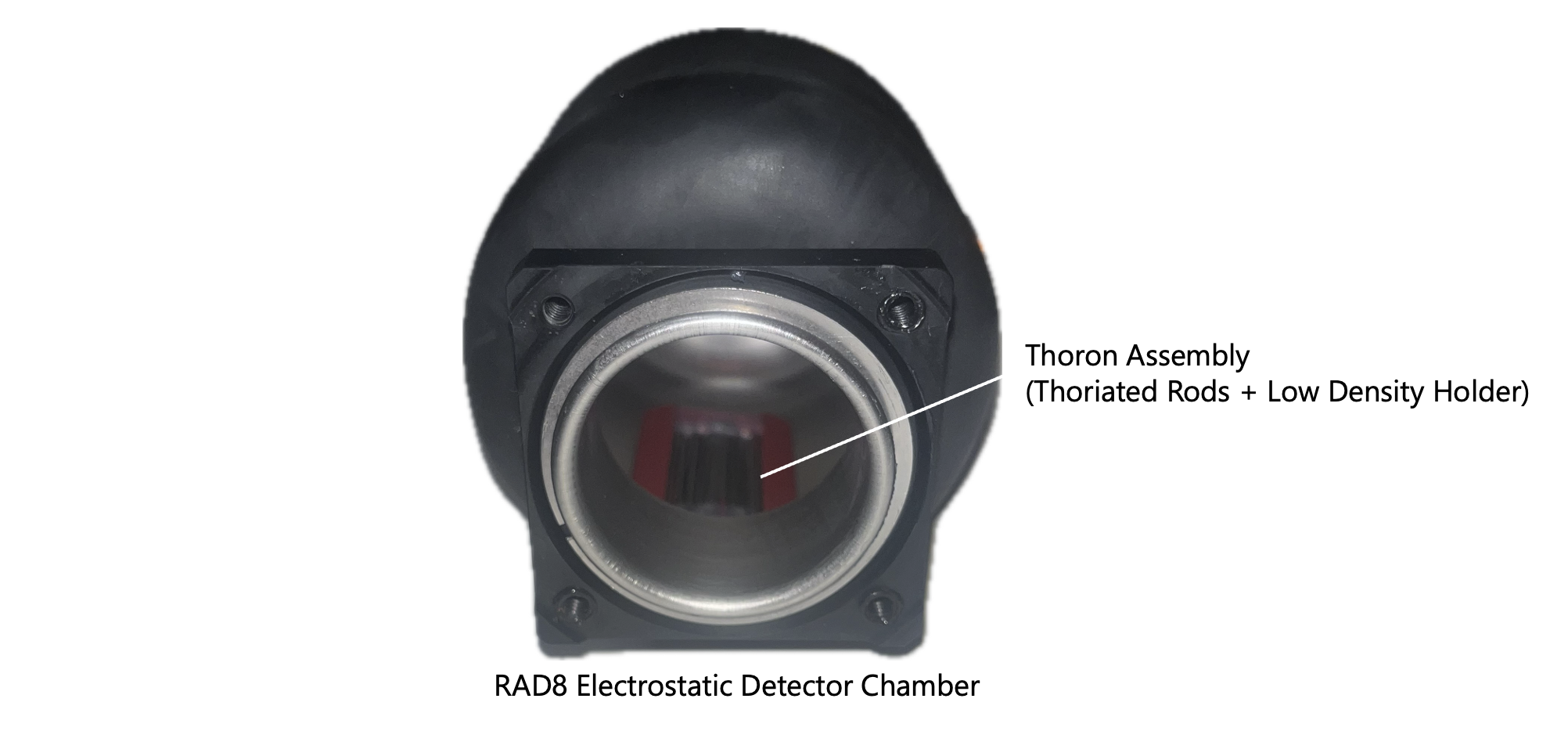} 
\caption{Photograph of the thoron assembly inside the RAD8 detector chamber.}
\label{fig:RAD8MECH_internal_photo}
\end{figure}

\autoref{fig:RAD8MECH_internal_photo} shows the assembly inside the RAD8 electrostatic detector chamber while it is detached from the instrument. The chamber is a \(0.6~\text{L}\) cylindrical volume with rounded edges and a port that aligns with the PIPS alpha detector when attached to the RAD8. The thoron assembly was inserted through this port and held in place by its own weight, with the 3D printed holder shaped to match the chamber curvature so that it remained aligned with the detector geometry. The chamber was then reattached to the RAD8, and the leak rate was verified to be consistent with the manufacturer specification.

A run duration of 3~hours was used to match the flowthrough measurements. Before each run, the detector chamber was flushed with low-humidity air until the RAD8 internal relative humidity readout was below 10\%. The chamber was then sealed and filled to 1~bar absolute pressure with ambient air. To provide an unambiguous ${}^{220}\mathrm{Rn}$ signature, the assembly was left to emanate for 12~hours before the start of each measurement run, allowing the build-up of later \(^{220}\mathrm{Rn}\) progeny. The \(^{216}\mathrm{Po}\) count rate and equilibrium intensity in Window B used for the ${}^{220}\mathrm{Rn}$ activity calculation is unchanged by the extended emanation period. Measurement runs were performed both with the thoron assembly inside the chamber and with the detector chamber empty, providing source and background data respectively.

\subsection*{Data analysis and results}
\begin{figure}[h]
  \centering
  \includegraphics[height=6.5cm]{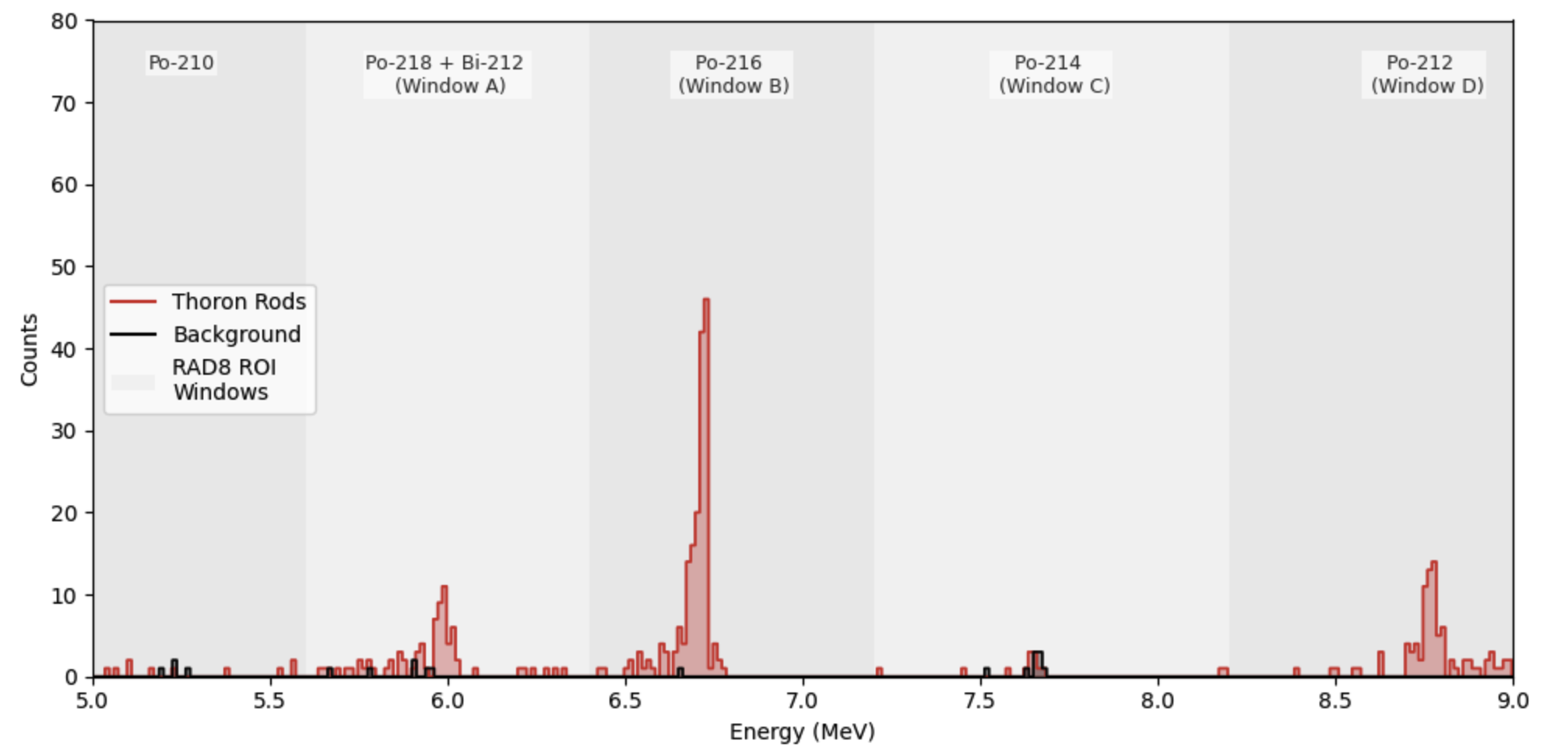}
\caption{Alpha energy spectra for the thoron source (red) and empty detector chamber background (black), measured with the in-chamber method in air over a 3~hour run.}
  \label{fig:internalmethodraw}
  
\end{figure}

\autoref{fig:internalmethodraw} shows the alpha energy spectrum for the thoron assembly and an empty detector chamber background using the in-chamber method, shown in red and black, respectively. The relative humidity in the setup remained below 15\% throughout both runs. A clear excess above background is observed in Window~B, consistent with \(^{216}\mathrm{Po}\). Excess counts are also visible in Windows~A and~D, consistent with \(^{212}\mathrm{Bi}\) and \(^{212}\mathrm{Po}\), providing an unambiguous ${}^{220}\mathrm{Rn}$ signature. 
The background-subtracted Window~B count rate was \((0.98 \pm 0.15)~\text{cpm}\), compared with \((0.32 \pm 0.09)~\text{cpm}\) from the flowthrough method for the same source \((A_{220} = 76 \pm 20~\mathrm{mBq})\) and the same run duration. This corresponds to a sensitivity improvement factor of \(3.1 \pm 1.0\) for the in-chamber method, indicating that placing the sample directly inside the detector chamber significantly reduces transfer losses.

\section{Enhancing sensitivity with helium}
\label{sec:helium}
\begin{figure}[h]
  \centering
  \includegraphics[height=6.5cm]{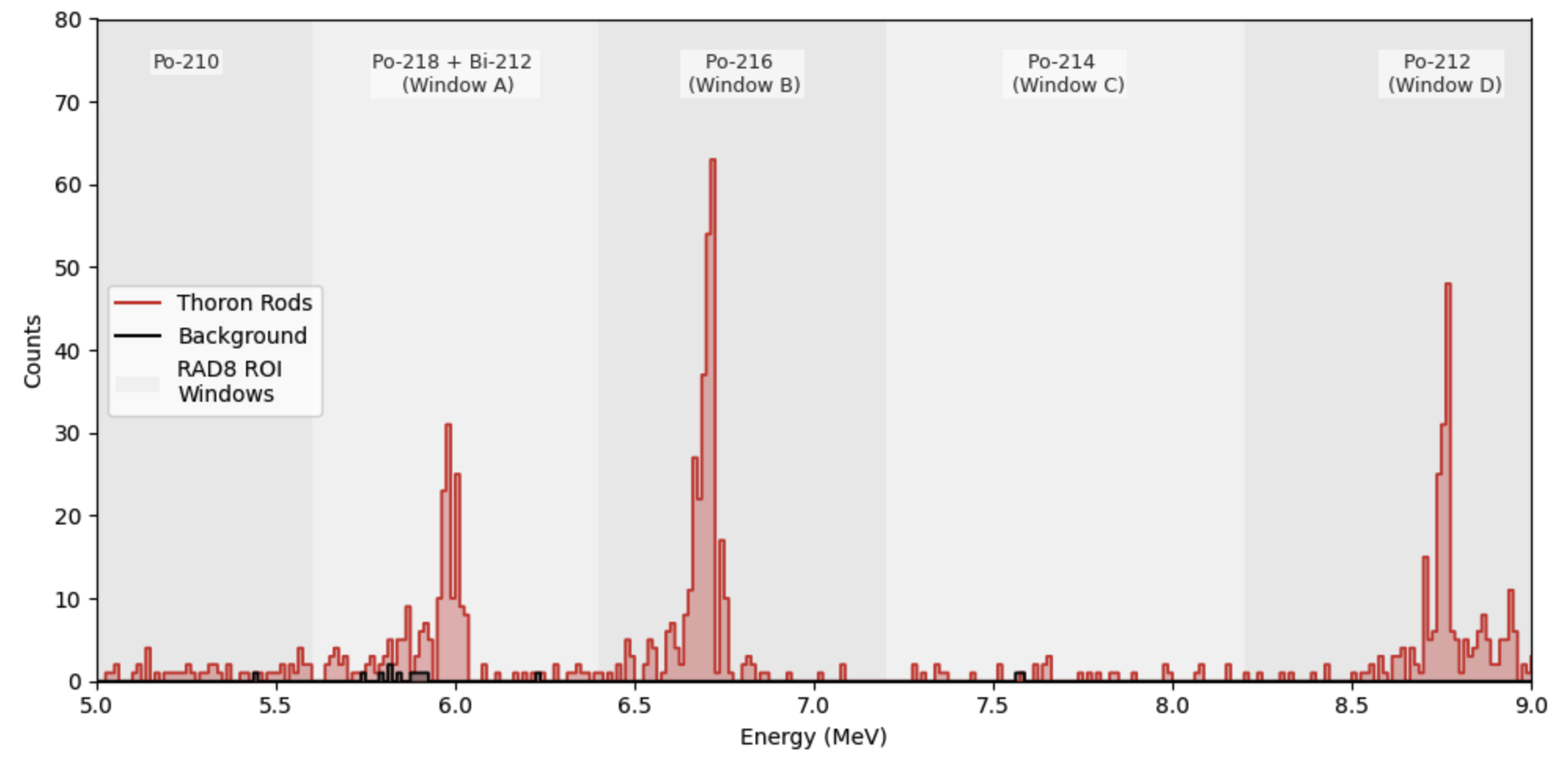}
\caption{Alpha energy spectra for the thoron source (red) and empty detector chamber background (black), measured with the in-chamber method in helium over a 3~hour run.}
  \label{fig:helium}
\end{figure}

Helium is known to improve the collection efficiency of radon in electrostatic detectors~\cite{wojcik2008high}. To further increase sensitivity, the in-chamber method described in \autoref{sec:inchamber} was repeated with 99.99\% grade helium as the fill gas instead of ambient air.
\autoref{fig:helium} shows the alpha energy spectrum integrated over a 3~hour measurement for the thoron assembly and an empty detector chamber background using the in-chamber method with helium, shown in red and black, respectively. There is a clear excess above background at the energies expected for ${}^{220}\mathrm{Rn}$ progeny, confirming a ${}^{220}\mathrm{Rn}$ signature in helium.

The background-subtracted Window~B count rate was \((1.71 \pm 0.19)~\text{cpm}\). Compared with the in-chamber measurement in ambient air, this corresponds to a sensitivity increase factor of \(1.7 \pm 0.3\) when using helium.  Relative to the flowthrough method, the overall sensitivity gain is \(5.3 \pm 1.6\). These results show that the in-chamber configuration with helium carrier gas further increases sensitivity.

\section{Discussions}
\label{sec:application}

In this section, the practical application of the in-chamber method is discussed, including its use in low-background material studies and the calibration to provide absolute \({}^{220}\mathrm{Rn}\) activity.

\paragraph{Detector chamber contamination}
A natural concern with placing samples directly inside the detector chamber is the risk of contaminating the chamber and affecting subsequent measurements. To assess this, background measurements with the in-chamber method were repeated after removing the thoron assembly from the detector chamber. The count rates in Window~B, which is the only window used to determine the ${}^{220}\mathrm{Rn}$ activity, were consistent within their uncertainties before and after the ${}^{220}\mathrm{Rn}$ runs, indicating no measurable residual contamination. Since Window~B is dominated by ${}^{216}\mathrm{Po}$ (half-life $0.145~\mathrm{s}$), its contribution decays back to baseline within a few minutes once the sample is removed and the chamber is flushed. This allows repeated in-chamber ${}^{220}\mathrm{Rn}$ measurements with only a short flush period between runs, supporting high throughput surveys.

\paragraph{Sensitivity and sample geometry}
\autoref{tab:sensitivity} summarises the Window~B count rates from the ${}^{220}\mathrm{Rn}$ assembly, which has a calibrated activity of \(76 ~\pm~ 20~\mathrm{mBq}\), for 3 hour runs using both the flowthrough and in-chamber methods. The observed sensitivity gain is expected to extend to other samples, provided they can be positioned within the low-density in-chamber holder, which has a cavity of approximately \(20 \times 50~\mathrm{mm}^2\) and thickness \(5~\mathrm{mm}\). Such constraints are typical of other low-background assay setups, where samples are prepared as coupons to fit a standard geometry~\cite{nguyen2025enhancing}.\\

\begin{table}[h]
\centering
\begin{tabular}{@{}lccc@{}}
\toprule
\textbf{Method}     & \textbf{Carrier gas} & \textbf{Window B (CPM)} & \textbf{Relative sensitivity} \\ 
\midrule
Flowthrough         & \ce{Air}             & \(0.32 \pm 0.09\)      & 1                     \\
In-chamber          & \ce{Air}             & \(0.98 \pm 0.15\)      & \(3.1 \pm 1.0\)               \\
In-chamber          & \ce{He}              & \(1.71 \pm 0.19\)      & \(5.3 \pm 1.6\)               \\
\bottomrule
\end{tabular}
\caption{Window~B count rates for the thoron assembly for different measurement methods and carrier gases.}
\label{tab:sensitivity}
\end{table}

\paragraph{Reduction factors and surface studies}
For many applications, such as the assessment of surface treatments and coatings, the key observable is the reduction factor in emanation rather than the absolute activity. In this use case, \({}^{220}\mathrm{Rn}\) emanation measured with the in-chamber method can be used to compare treated and untreated samples directly, since the absolute detection efficiency largely cancels in the ratio due to minimal change to the overall geometry of the sample in these treatments. Moreover, in many dark matter detectors operated at cryogenic temperatures, the dominant $^{222}\mathrm{Rn}$ contribution arises from recoil-induced emanation from surfaces and near-surface regions rather than diffusion from the bulk. In this context, in-chamber $^{220}\mathrm{Rn}$ measurements can serve as a rapid first-pass proxy for $^{222}\mathrm{Rn}$ surface behaviour. Surface studies that would normally require several weeks of $^{222}\mathrm{Rn}$ accumulation can instead be carried out on timescales of hours. This enables rapid pre-screening of treatments and materials so that only the most promising candidates are followed up with high-sensitivity $^{222}\mathrm{Rn}$ measurements that require longer accumulation times.

\subsection*{Absolute calibration of the in-chamber method}
If absolute ${}^{220}\mathrm{Rn}$ emanation rates are required, it is first necessary to quantify how placing the sample inside the detector chamber modifies the collection efficiency. This can be achieved by utilising the RAD8 response, which separates the ${}^{220}\mathrm{Rn}$ and ${}^{222}\mathrm{Rn}$ decay chains into distinct analysis windows.

In the in-chamber setup, ${}^{220}\mathrm{Rn}$ is produced directly in the active detector volume. The progeny would therefore be expected to be collected with the same efficiency as those from ${}^{222}\mathrm{Rn}$, for which transfer losses are negligible. However, introducing the thoron assembly into the detector chamber distorts the electric field and reduces the effective collection efficiency. Once this suppression is quantified, the resulting effective collection efficiency can be used to infer the ${}^{220}\mathrm{Rn}$ activity.

The suppression can be experimentally determined by measuring a steady \({}^{222}\mathrm{Rn}\) source with and without the thoron assembly in the detector chamber. \({}^{222}\mathrm{Rn}\) produces signal in Windows~A and~C from the \({}^{218}\mathrm{Po}\) and \({}^{214}\mathrm{Po}\) peaks, respectively. There is also a contribution from \({}^{220}\mathrm{Rn}\) to Window~A via \({}^{212}\mathrm{Bi}\), but, under the conditions considered here, Window~C receives counts only from the \({}^{222}\mathrm{Rn}\) decay chain. Since the thoron assembly does not produce a measurable signal in Window~C above background, the ratio of the Window~C rates with and without the assembly provides a direct measure of the suppression of the collection efficiency. This suppression factor can then be used to infer the effective collection efficiency for \({}^{220}\mathrm{Rn}\) with the thoron assembly in the detector chamber.

\begin{figure}[h]
  \centering
  \includegraphics[height=5cm]{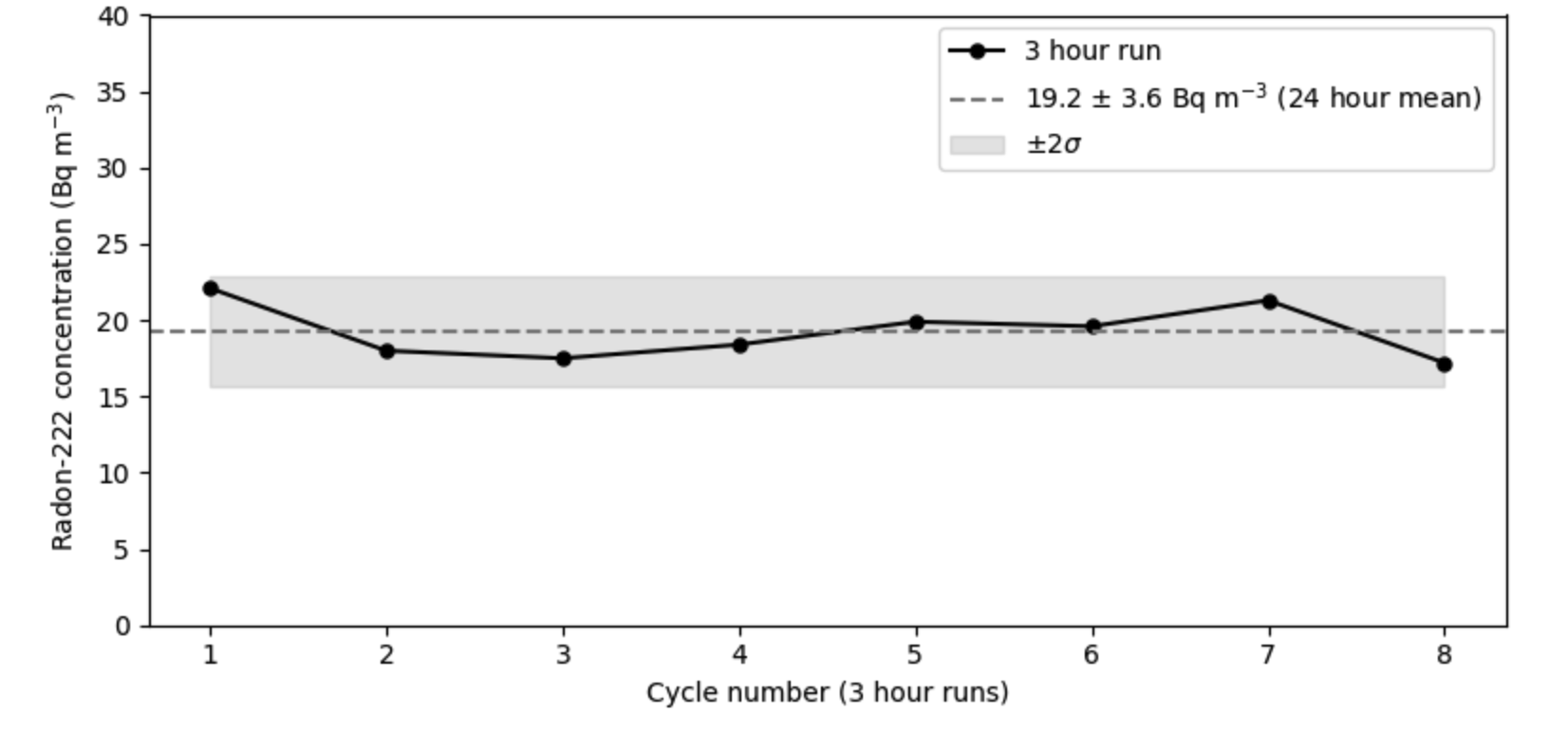}
  \caption{Plot of concentration over a 24-hour reference run with low activity \({}^{222}\mathrm{Rn}\) source.}
  \label{fig:radon_stability}
\end{figure}

\autoref{fig:radon_stability} shows the compressed air flowthrough low-activity \({}^{222}\mathrm{Rn}\) source used for the suppression calibration. The RAD8 with an empty chamber was run in 3 hour cycles, to mimic the ${}^{220}\mathrm{Rn}$ measurement duration, for a total of 24~hours. The resulting  ${}^{222}\mathrm{Rn}$ activity averaged \(19.2 \pm 3.6~\mathrm{Bq\,m^{-3}}\), which corresponds to a Window~C rate of \(0.23 \pm 0.03~\mathrm{cpm}\). The 24-hour measurement was then repeated with the thoron assembly inserted in the detector chamber, resulting in a Window~C rate of \(0.19 \pm 0.02~\mathrm{cpm}\).

The absolute ${}^{220}\mathrm{Rn}$ activity \(A_{220}\) can then be extrapolated using
\begin{equation}
  A_{220}
  \;=\;
  \frac{R_B^{\text{net}}\,V_{\text{reduced}}}
       {\left(R_C^{\text{assembly}}/R_C^{\text{empty}}\right)\,k_{\text{air}}}\,,
  \label{eq:A220_calib}
\end{equation}
where \(R_C^{\text{assembly}}/R_C^{\text{empty}} = 0.84 \pm 0.14\) ~is the suppression factor, 
\(k_{\text{air}} = 0.0104~\mathrm{cpm}\,(\mathrm{Bq\,m^{-3}})^{-1}\) is the unsuppressed sensitivity from the manufacturer calibration in air, 
\(R_B^{\text{net}}\) is the background-subtracted Window~B count rate for the ${}^{220}\mathrm{Rn}$ run of interest, 
and \(V_{\text{reduced}} = 0.59~\mathrm{L}\) is the detector chamber volume minus the volume of the sample. Using the in-chamber ${}^{220}\mathrm{Rn}$ run in air from \autoref{tab:sensitivity} gives
\[
  A_{220} = 69 \pm 19~\mathrm{mBq}\ (95\% ~\text{C.L.})\,.
\]
This is consistent, within uncertainties, with the independently measured value of \(76 \pm 20~\mathrm{mBq}\) calculated using the conventional flowthrough setup, demonstrating that the in-chamber technique can be used to measure absolute \({}^{220}\mathrm{Rn}\) activity  as well as relative reduction factors.

Although this work has focused on ${}^{220}\mathrm{Rn}$, the in-chamber concept can in principle be adapted to ${}^{222}\mathrm{Rn}$ emanation measurements. Since leading radon emanation facilities often rely on large external emanation chambers and complex enrichment to achieve high sensitivity, intrinsic emanation from the chambers and connecting pipework can become a significant background contribution. In cases where this is the leading limitation on the measurement sensitivity, an in-chamber ${}^{222}\mathrm{Rn}$ configuration would eliminate separate emanation volumes, thereby reducing the total emanating surface area to that of the detector chamber alone and removing transfer inefficiencies associated with complex enrichment and transfer lines.


\section{Conclusions}
\label{sec:conclusions}

In this paper, a direct in-chamber method for measuring ${}^{220}\mathrm{Rn}$ emanation with a DURRIDGE RAD8 electrostatic detector has been presented. An industry-standard flowthrough emanation setup was used as a baseline to measure the low-activity thoron source at $76 \pm 20~\mathrm{mBq}$. For the same source, the in-chamber configuration improved the sensitivity by a factor $\sim$3 in air, increasing to a factor $\sim$5 when helium was used as the carrier gas. An absolute calibration based on the ${}^{222}\mathrm{Rn}$ response in Window~C gave $A_{220} = 69 \pm 19~\mathrm{mBq}$ (95\% C.L.), consistent with the independently calibrated flowthrough value of $76 \pm 20~\mathrm{mBq}$ and showing that the in-chamber technique can provide enhanced sensitivity for both reduction-factor 
measurements and absolute activity measurements. These findings demonstrate in-chamber ${}^{220}\mathrm{Rn}$ emanation measurements as a practical tool for low-background material screening and radon mitigation studies, enabling rapid pre-screening of surface treatments by using ${}^{220}\mathrm{Rn}$ as a short-lived proxy for ${}^{222}\mathrm{Rn}$ surface emanation. The in-chamber technique could be adapted to ${}^{222}\mathrm{Rn}$ to reduce intrinsic emanation and support targeted high-sensitivity studies in next-generation rare-event experiments.

\acknowledgments
The authors would like to acknowledge support for this work through the Australian Research Council Centre of Excellence for Dark Matter Particle Physics Grant (CE200100008).




\bibliographystyle{JHEP}  
\bibliography{references}  

\end{document}